# Monopole condensation and dual superconductivity: the $SU(2)$ case.

L. Del Debbio, A. Di Giacomo, G. Paffuti and P. Pieri * a.

aDipartimento di Fisica dell'Università and I.N.F.N.,
Piazza Torricelli 2, I-56100 Pisa, Italy.

We study the behaviour of a suitably defined disorder parameter, showing for the first time monopole condensation in the ground state of $QCD$.

## 1. Introduction

Dual superconductivity has been often advocated as the mechanism responsible for confinement. According to this scenario, magnetic charges, defined as Dirac monopoles of a residual $U(1)$ symmetry selected by a suitable gauge fixing, should condense to produce superconductivity. Although a clear evidence has been provided that monopole do exist and may play a role at the deconfinement phase transition of lattice gauge theories [1], a proof of monopole condensation is still lacking. Monopole condensation means that the ground state of the theory is a superposition of states with different magnetic charges, so that the dual (magnetic) $U(1)$ symmetry is broken, in complete analogy with the electric $U(1)$ symmetry breaking in an ordinary superconductor. Such a breaking is monitored by any operator with non-trivial magnetic charge, which should exhibit a non-vanishing vacuum expectation value. The detailed construction of a disorder parameter has been presented elsewhere at this conference [2]. We generalize this construction for monopoles defined in non-Abelian pure gauge theories and use our operator to prove the vacuum structure of these theories.

## 2. Monopoles in pure $QCD$

It is a well-known result that stable monopole solutions are related to the elements of the first homotopy group of the gauge group. Since for $SU(N)$, we have:

$$\pi_1 (SU (N)) = \{1\} \tag{1}$$

the gauge symmetry has to break down to some non simply connected subgroup, in order to define magnetic charges. In the Georgi & Glashow model, the presence of a matter (scalar) field, minimally coupled to the gauge field, producing a spontaneous symmetry breaking

$$SU(2) \to U(1),$$

allows to define an Abelian field strength tensor:

$$f_{\mu\nu} = G_{\mu\nu}^a \phi^a - \frac{1}{g}\varepsilon^{abc}\hat{\phi}^a D_\mu \hat{\phi}^b D_\nu \hat{\phi}^c \tag{2}$$

which admits monopole solutions (the 't Hooft-Polyakov monopoles).

One can perform a gauge rotation such that

$$\hat{\phi}^a = \delta^{a3}. \tag{3}$$

In this gauge:

$$f_{\mu\nu} = \partial_\mu a_\nu - \partial_\nu a_\mu \tag{4}$$

where $a_\mu = A_\mu^a \hat{\phi}^a$. Eq. (4) is the usual expression for the $U(1)$ electro-magnetic field built from the Abelian potential $a_\mu$. The 't Hooft-Polyakov solution for a monopole located in the origin is simply given by the Dirac potential $b_i$:

$$a_i = \varepsilon_{3ik}\frac{x_k}{r(r - x_3)} \equiv b_i \tag{5}$$

if we put the string along the positive $x_3$-axis. The monopole creation operator we introduced

*Partially supported by MURST and by EC Contract CHEX-CT92-0051

for the $U(1)$ case in [2] can be easily generalized in this gauge: we simply need an operator shifting the $a$ field by an amount $b$. The result is:

$$\mu(x;b) = \exp\left\{i\int d^3\mathbf{y}\; f_{0i}(\mathbf{y},x_0)b_i(\mathbf{y}-\mathbf{x})\right\} \quad (6)$$

Notice that $\mu(x;b)$ is invariant under $SU(2)$ gauge transformations.

In pure gluodynamics, the residual symmetry which allows to define monopoles is selected by Abelian Projection ($AP$). There are essentially two ways to perform $AP$:

1. diagonalizing an operator $X$ which transforms in the adjoint representation under gauge transformations, we will call this choices 't Hooft gauges;

2. maximizing a given operator (maximal Abelian gauge).

In the first case, writing $X$ as

$$X(x) = \exp\left\{i\xi^a(x)\sigma^a\right\}, \quad (7)$$

a field $\xi$ playing the role of the $\phi$ field of the 't Hooft-Polyakov monopole, and hence the $f_{\mu\nu}$ tensor can be defined, using Eq. (2). On the other hand, in the maximal abelian gauge, we have no recipe to identify such a field. The only way to construct $f_{\mu\nu}$ is to perform $AP$, to extract the abelian part of each link, $a_\mu$, and evaluate $f_{\mu\nu} = \partial_\mu a_\nu - \partial_\nu a_\mu$ for each update. This procedure is extremely time-consuming and happens to be the major problem when we try to extend our analysis to the maximal Abelian gauge.

## 3. Monopole creation operator

On the lattice our operator is the *naïve* translation of the continuum quantity written above:

$$\mu^L(x;b) = \exp\left\{-\beta\sum f_{0i}(\mathbf{y},x_0)b_i(\mathbf{y}-\mathbf{x})\right\}$$
$$\equiv \exp\{-\beta S_b\}$$

where $f_{0i}$ is given by the lattice version of Eq. (2). As in the $U(1)$ case, we define the disorder parameter as:

$$\langle\bar\mu\rangle = \frac{\langle\mu\rangle}{\langle\gamma\rangle}$$

where

$$\gamma(x;g) = \exp\left\{-\beta\sum f_{0i}(\mathbf{y},x_0)g_i(\mathbf{y}-\mathbf{x})\right\}$$
$$\equiv \exp\{-\beta S_g\}$$

and $g$ is a constant field properly normalized (see [2,3], for details).

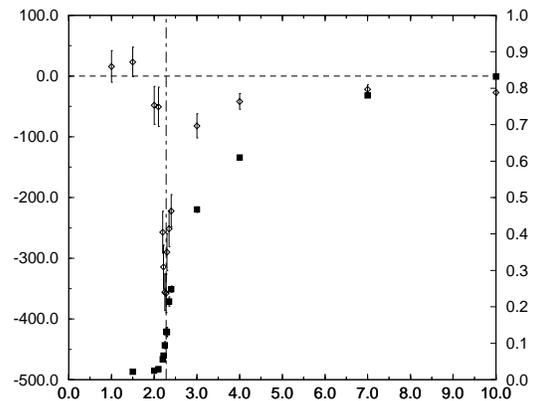

Figure 1. The $\rho$ operator on a $12^3 \times 4$ lattice.

## 4. Numerical results

As in the $U(1)$ case [3], we define

$$\rho = \frac{d}{d\beta}\ln\langle\bar\mu\rangle$$

We tested two different types of monopoles defined à la 't Hooft, using as $X$ operator first the Polyakov line, then the $(1,2)$ plaquette.

The results for the Polyakov gauge are reported in the following figures. In fig. 1, we have the behaviour of $\rho$ vs. $\beta$ (diamonds), on a $12^3 \times 4$ lattice, together with the Poliakov line (black squares). We see that $\rho$ shows a clear peak at $\beta_c$, exactly at the phase transition, where the Polyakov line has a drop. This result provides the first evidence that monopole condensation indeed occurs. In order to be sure that the peak we are seeing is really related to monopoles, we repeat the same measurements with different external potentials (the $b_i$ in (6)): only for those potentials which has a monopole topology, we recover some signal at the phase transition). Moreover, on lattices with a different temporal size, the critical coupling $\beta_c$ corresponding to the phase transition changes in order to keep the physical critical temperature constant, according to:

$$T^{\text{phys}} = \frac{1}{N_t a(\beta)}$$

The peak in our operator shifts together with $\beta_c$. When growing the volume of our lattices, we note, even at a very qualitative level, that the peak we have obtained gets deeper and narrower, as expected since $\rho$ should become a $\delta$-function in the infinite volume limit. Comparing the results for lattices of sizes $8^3 \times 4$, $12^3 \times 6$ and $16^3 \times 8$, which all have the same physical size, we see that the height of the peak do not change significantly. We take this as an indication that $\rho$ scales with the physical volume.

Instead of $\bar{\mu}$ we can study the correlation between a monopole and an anti-monopole operator:

$$C(d) = \langle \mu(x; b)\mu(x'; \bar{b})\rangle \tag{8}$$

Here the total charge of the system is zero and we don't have to worry about boundary conditions. We expect the cluster property to hold for large distances:

$$C(d) \xrightarrow{d \to \infty} \langle \mu \rangle^2 \tag{9}$$

which means:

$$\rho_{(b,\bar{b})}(d) \equiv \frac{d}{d\beta}\ln\frac{C(d)}{\langle\gamma\rangle^2} \xrightarrow{d \to \infty} 2\,\rho \tag{10}$$

The data for the correlation are reported in Fig. 2: circles and triangles represent respectively the values of $\rho_{(b,\bar{b})}$ and $2\rho$. The difference between them is given by the black squares and is always compatible with zero.

As in the Abelian case, we checked for the effect the independence of boundary conditions: the result is positive and will be reported in greater detail elsewhere [3].

Finally it is interesting to note that we find no signal of condensation if we define the monopoles in the gauge selected diagonalizing the plaquette. Our data do not show a clear peak, which seems to indicate that not all $AP$ define monopoles relevant to confinement.

## 5. Conclusions

We can briefly summarize our results as follows:

- we have defined a disorder parameter monitoring condensation monopoles defined by different $AP$ and have succesfully applied it, getting an evidence that monopoles do condense in the ground sate of $QCD$;

- we have performed some "consistency" checks on our operator (finite size efffects, cluster property, boundary conditions, other topologies).

- we have found that monopoles defined by different $AP$ are not equivalent. We are studying other gauge fixings, in particular we are trying to extend our construction to the maximal abelian gauge.

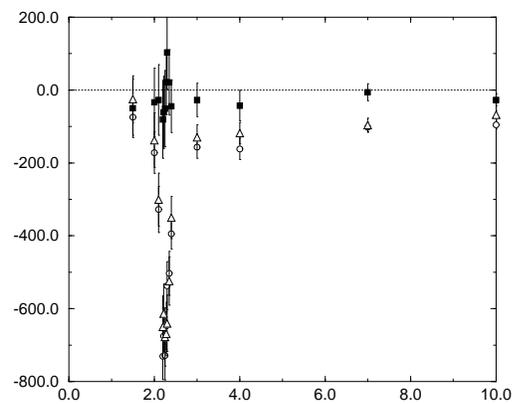

Figure 2. The correlation on a $12^3 \times 4$ lattice.